\documentclass[aps,pre,reprint]{revtex4-1}
\usepackage{amsmath}
\usepackage{graphicx}
\usepackage{color}
\usepackage{amssymb}
\usepackage{wasysym}
\usepackage{soul}
\input{epsf}

\begin{document}

\bibliographystyle{prsty}

\title{Heterogeneous Dynamics in Columnar Liquid Crystals of Parallel Hard Rods}

\author{Simone Belli}
\email{s.belli@uu.nl}
\affiliation{Institute for Theoretical Physics, Utrecht University, Leuvenlaan 4, 3584 CE, Utrecht, The Netherlands}
\author{Alessandro Patti}
\affiliation{Soft Condensed Matter Group, Debye Institute for NanoMaterials Science, Utrecht University, Princetonplein 5, 3584 CC, Utrecht, The Netherlands}
\author{Ren\'{e} van Roij}
\affiliation{Institute for Theoretical Physics, Utrecht University, Leuvenlaan 4, 3584 CE, Utrecht, The Netherlands}
\author{Marjolein Dijkstra}
\email{m.dijkstra1@uu.nl}
\affiliation{Soft Condensed Matter Group, Debye Institute for NanoMaterials Science, Utrecht University, Princetonplein 5, 3584 CC, Utrecht, The Netherlands}

\begin{abstract}

In the wake of previous studies on the rattling-and-jumping diffusion in smectic liquid crystal phases of colloidal rods, we analyze here for the first time the heterogeneous dynamics in columnar phases. More specifically, we perform computer simulations to investigate the relaxation dynamics of a binary mixture of perfectly aligned hard spherocylinders. We detect that the columnar arrangement of the system produces free-energy barriers the particles should overcome to jump from one column to another, thus determining a hopping-type diffusion. This phenomenon accounts for the non-Gaussian inter-column diffusion and shows a two-step structural relaxation which is remarkably analogous to that of out-of-equilibrium glass-forming systems and gels. Surprisingly enough, slight deviations from the behavior of simple liquids due to transient cages is also observed in the direction perpendicular to this plane, where the system is usually referred to as liquid-like.
\end{abstract}

\maketitle

\section{Introduction}
\label{sec:Introduction}

Liquid crystals (LCs) are phases of matter in which the anisotropy of the particles determines under specific conditions a partial spontaneous breaking of the spatial symmetries of the system, thus manifesting features in between the crystalline solid and the isotropic liquid phase. The notion that entropic effects alone are sufficient to drive the self-assembly of ordered liquid crystal phases is well established in colloid science \cite{onsager, israelachvili, forsyth, flory, frenkel2}. Model systems of hard particles constitute the natural choice to describe the phase and aggregation behavior of most colloidal systems, as the main interactions established between their particles have a repulsive, steric origin. As a consequence, a hard-particle fluid does not have internal energy and minimizing its free energy is equivalent to maximizing its entropy.

In his seminal work, Onsager showed that mere hard-core repulsions between infinitely thin rigid cylinders are able to determine an entropy-driven phase transition from the isotropic to the nematic phase, and hence the existence of a spontaneous orientational order \cite{onsager}. The evolution of simulation techniques in the past decades allowed to investigate the more realistic case of rods with finite size, showing that by varying the length-to-diameter ratio one- (smectic), two- (columnar) and three- (crystal) dimensional translational ordered phases can be encountered \cite{stroobants1,stroobants2}. Further more accurate studies showed that for a monodisperse system of both aligned \cite{veerman} and freely rotating \cite{bolhuis} hard spherocylinders the columnar phase happens to be metastable with respect to the smectic phase for each value of the length-to-diameter ratio. The complexity of the phase behavior of linear particles becomes even more pronounced by proceeding from monodisperse systems to mixtures \cite{stroobants,cui,bates,martinez,varga}. In particular, size polydispersity introduces a sensible change in the phase behavior of rod-like particles as their packing is not as effective as that of monodisperse rods. In fact, it was observed that in a system of hard spherocylinders the formation of smectic layers can be inhibited by introducing a length bidispersity, in such a way that the columnar phase can become thermodynamically stable \cite{stroobants,cui}. Entropy-driven columnar phase transitions have been observed in monodisperse systems of disc-like particles, such as cut spheres or oblate spherocylinders \cite{frenkel3,veerman2,cuetos,filion}. As far as rod-like particles are concerned, theoretical studies indicated that the columnar order can be observed not only in bidisperse mixtures, but also in more realistic polydisperse systems of parallel cylinders \cite{bohle}. Polydispersity is not the only element which favors the stabilization of the columnar phase in a system of rods, as was shown in Ref. \cite{wensink} where a monodisperse system of soft-core rods was considered. On the other hand, the effect of rod flexibility in stabilizing the columnar phase is still under debate \cite{hentschke,sear,cinacchi,grelet2}.

With the improvement in understanding and describing static and thermodynamic properties of LCs, the interest towards the dynamics correspondingly increased. In particular, most of the studies in this direction were devoted to analyze the anisotropy of the diffusion by the measurement of the self-diffusion coefficients in different mesophases \cite{hess,lowen,bates2,cifelli}. For lyotropic LCs these results found good overall agreement with experiments based on techniques such as fluorescence recovery after photobleaching (FRAP) \cite{bu,vbruggen1,vbruggen2,cush}. On the other hand, only few studies focused on the analysis of the dynamical phenomena at the single-particle level, where fluorescence microscopy was applied to investigate the LC phases in colloidal suspensions of \textit{fd} virus \cite{lettinga2,lettinga}. This approach allowed to observe for the first time the mechanism of interlayer diffusion, or permeation, which characterizes the dynamics of the system in the smectic phase \cite{lettinga}. In fact, the layered structure of smectic LCs, which determines an effective periodic mean-field potential, influences the motion in the direction perpendicular to the smectic layers with the appearance of jumps of the order of the rod length. As a result of this quasi-quantized diffusion, at intermediate time intervals one can distinguish between ``slow'' particles, which rattle around the center of a column, and ``fast'' particles, which jump to another column. This heterogeneous dynamics accounts for deviations from Gaussianity in the diffusion, a phenomenon which is also well known in homogeneous complex fluids such as supercooled liquids \cite{kob,gray} and gels \cite{puertas,gao}. In the light of these results, a theoretical approach based on dynamic density functional theory using the second virial approximation focused on the role played by the local fluid structure of the system, which competes with the permanent barriers due to the smectic structure, thus increasing even more the analogies with fluids close to a dynamical arrest transition \cite{bier,grelet}. Simulations on aligned \cite{matena} and freely rotating \cite{patti,patti2} hard spherocylinders confirmed qualitatively both these studies and pointed out the effect of dynamical heterogeneities on the structural relaxation of the system, which deviates from the exponential decay expected for simple fluids. Furthermore, in Refs. \cite{patti,patti2} it was shown that the heterogeneous dynamics in the smectic phase of hard spherocylinders is tightly related to cooperative motion of particles through the smectic layers.

Following this line of research and motivated by a recent experiment on \textit{fd} virus \cite{pouget}, we investigate in this paper the dynamics of a binary mixture of rod-like particles which exhibits a stable columnar liquid crystal phase. Using Monte Carlo (MC) simulations we are able to study for the first time the dynamical heterogeneities which arise from the columnar structure and their effect on the diffusion and on the long-time structural relaxation of the system. Furthermore, we measure the height of the inter-column energetic barriers and compare our results with the simulation data available for the smectic phase \cite{matena,patti,patti2}. Conventional MC dynamics has been extensively applied to study the origin of non-exponential relaxation of glass-forming liquids \cite{doliwa,doliwa2,berthier,brambilla,berthier2}. MC and Molecular Dynamics (MD) simulations of Lennard-Jones fluids \cite{huitema} showed that the particle dynamics at large time scales are the same, but are different at small time scales, as the stochastic motion of particles cannot be detected by the deterministic approach of MD simulations.

The paper is organized as follows. In section II we introduce the model, the simulation details and the physical properties which were measured in order to describe the dynamics of the system. The results of these measurements, which confirm the presence of dynamical heterogeneities also in the columnar phase and are in general agreement with the observations in the smectic phase, are discussed in section III, whereas in section IV we illustrate our conclusions.

\section{Model and Simulations}
\label{sec:Model and Simulations}

We study a system containing $N=1600$ perfectly aligned hard spherocylinders with aspect ratio $L^{*}=L/D$, where $L$ and $D$ are, respectively, the length and diameter of a cylindrical body capped by two hemispheres with diameter $D$. The phase diagram of a monodisperse system containing such rod-like particles shows stable nematic, smectic, and crystal phases, but lacks a stable columnar phase in the range $0\leq L^{*}< \infty$ \cite{veerman}. Stroobants studied the phase behavior of bidisperse systems of hard rods and found that the bidispersity can favor and stabilize columnar order over smectic order \cite{stroobants}. Therefore, to prevent the formation of smectic layers, we investigate a binary mixture of hard spherocylinders with the same diameter $D$ (used as our unit of length), but different lengths $L^{*}_{1}$ and $L^{*}_{2}$, with $L^{*}_{1}>L^{*}_{2}$. In this model, the rotational degrees of freedom are frozen out and hence the particles are forced to be aligned along a common nematic director, oriented along the \textit{z} axis. The relative concentration of the two species is set in such a way that the binary mixture is kept at its equivalence point, where the volume fractions of each component are the same. The phase diagram at fixed $L^{*}_{2}=1.0$ displays a region of stability of the columnar phase which increases with $L^{*}_{1}$ and disappears at $L^{*}_{1} \leq 1.6$, where a nematic-smectic transition is observed \cite{stroobants}. Here we study a columnar ordered binary mixture of rods with $L^{*}_{1}=2.1$ and $L^{*}_{2}=1.0$, and relative concentrations $x_{1}=N_{1}/N=0.375$ and $x_{2}=N_{2}/N=0.625$, respectively. For lower pressures, this columnar phase transforms into a nematic phase, while for higher pressures it freezes into a crystal phase.

We performed standard MC simulations in a rectangular box of volume $V$ with periodic boundary conditions. To equilibrate the columnar phase, we performed runs in the isobaric-isothermal (\textit{NPT}) ensemble, where the particle moves were accepted according to the Metropolis algorithm \cite{frenkel}, that is if no particle overlap was detected. Each MC cycle consisted of \textit{N} attempts to displace a randomly selected particle, plus an attempt to modify the box volume with independent changes of the three box sides. The system was considered to be in equilibrium when the volume reached a stationary value within the statistical fluctuations. We run simulations at several reduced pressures $P^{*}=\beta PD^{3}$, where $\beta = 1/k_{B}T$, $k_{B}$ is the Boltzmann's constant and \textit{T} the absolute temperature. In particular, we equilibrated a nematic phase at $P^{*}=2.5$ (packing fraction $\eta= N(x_1 v_1+x_2 v_2)/V=0.470$ with $v_i$ ($i=1,2$) the single particle volume) which is very close to the nematic-columnar transition, and three different columnar phases at $P^{*}=3.0$ ($\eta=0.535$), $3.5$ ($\eta=0.563$) and $4.0$ ($\eta=0.580$). In all these cases, our starting configuration consisted of a highly packed columnar structure with the rods randomly located along the \textit{z} direction, and hexagonally ordered in the \textit{xy} plane. The minimum number of MC cycles needed for an equilibration run was $5\times10^{5}$, and was followed by a production run of $2\times10^{6}$ MC cycles in the canonical (\textit{NVT}) ensemble to simulate the relaxation dynamics and evaluate all the physical properties of interest. In this case, the box volume was kept fixed to prevent unphysical collective moves which do not mimic the Brownian dynamics of the particles properly. It was proved that in rod suspensions the contribution of hydrodynamics can at first approximation be neglected with respect to steric effects, which result in excluded volume interactions \cite{pryamitsyn}. Under these conditions, the MC approach offers an important tool to study the dynamics of colloids, since, in spite of its intrinsically non-dynamical nature, it is able to reproduce the Brownian diffusion typical of such kind of systems \cite{sanz}. To pursue this goal, one must set a small enough maximum MC displacement, typically of the order of one tenth of the shortest dimension of the particle. The optimal value of the mean particle displacement is strictly linked to the acceptance rate and hence to the CPU time per simulation run. If the displacement is chosen too small, the system would need longer runs to properly explore the configurational space. The same effect is expected with too big displacements, as most of the particle moves would cause overlaps and therefore would be rejected. However, any reasonable and  convenient choice of the average step size should not affect the dynamics at long time scales. We tested that, apart from an overall scaling, this was the case. The maximum displacement was fixed to give an acceptance rate of roughly 50\% per move. Furthermore, to take into account the non-spherical shape of the particles, the maximum MC displacement was set in such a way that at short times it reproduces the anisotropic diffusion of a single colloidal rod. More specifically, this means that the ratio between the maximum MC displacement in the $xy$ plane $\Delta x_{max}=\Delta y_{max}$ and in the $z$ direction $\Delta z_{max}$ must be set as

\begin{equation}
\frac{\Delta x_{max}}{\Delta z_{max}}=\frac{\Delta y_{max}}{\Delta z_{max}}=\sqrt{\frac{D_{\perp}}{D_{\parallel}}},
\label{eq0} 
\end{equation}
where we denoted with $D_{\perp}$ and $D_{\parallel}$ the short-time self-diffusion coefficients of the rod in the direction parallel and perpendicular to its long axis, respectively. To give an estimate for the ratio $D_{\perp}/D_{\parallel}$, which solely depends on the geometry of the particle, we referred to the semi-empirical expression derived in Ref. \cite{tirado}. In that work, the authors used a numerical approach to evaluate the translational self-diffusion coefficients of a cylindrical particle for different values of the length-to-diameter ratio $p$, finding good overall agreement with experimental data in the range $2<p<30$ \cite{tirado1}. A least-square quadratic fitting in $p^{-1}$ of the data allowed the authors to give an expression for the transverse and longitudinal self-diffusion coefficients as functions of the parameter $p$. Since here we consider spherocylinders, we evaluated the ratio $D_{\perp}/D_{\parallel}$ by setting $p=(L+D)/D$, arguing that possible deviations given by the non-cylindrical shape of the particles were irrelevant. According to the above-mentioned expression, the ratio between the maximum MC displacement perpendicular and parallel to the $z$ axis was set to $0.92$ for particles of species $1$ and $0.94$ for those of species $2$. Moreover, since the transverse section of the particles is the same for the two species, we set the same maximum MC displacement along the $z$ axis for the two components. Once the short-time self-diffusion coefficients are known, it is possible to introduce a time scale defined by $\tau=D^{2}/D_{tr}$, where the total translational diffusion coefficient $D_{tr}=(\langle D_{\parallel} \rangle +2 \langle D_{\perp} \rangle)/3$ is evaluated in terms of the longitudinal and transverse short-time diffusion coefficients averaged over the two species \cite{matena}.

In order to analyze the heterogeneous diffusion and the structural relaxation of the system, the following physical properties were calculated: (i) the transverse mean-field potential, (ii) the self-part of the van Hove function (SVHF), (iii) the distinct-part of the van Hove function (DVHF), (iv) the mean square displacement (MSD), (v) the non-Gaussian parameter (NGP) and (vi) the self-part of the intermediate scattering function (SISF).

\textit{\mdseries{Transverse mean-field potential.}} In a liquid crystal phase characterized by columnar order the translational invariance is spontaneously broken in the plane perpendicular to the nematic director. This gives rise to a non homogeneous (relative) probability $\pi_{i}(x,y)$ of finding a particle of species $i=1,2$ at position $(x,y)$ in the plane perpendicular to the nematic director. The effective energetic barrier which tends to confine the particle inside a column is given by the mean-field potential $U_{i}(x,y)$, defined as \cite{lettinga}

\begin{equation} 
\pi_{i}(x,y) \propto \exp \left[ -\frac{U_{i}(x,y)}{k_{B} T} \right],
\label{eq1}
\end{equation}
where the proportionality constant is chosen in such a way that the minima of the potential is set to zero.

\textit{\mdseries{Self-part of the van Hove function.}} The heterogeneous dynamics and hopping-type inter-column diffusion can be quantitatively described by the SVHF \cite{vanhove}

\begin{equation} 
 G_{s}(\mathbf{r},t) = \frac{1}{N} \left\langle \sum_{j=1}^{N} \delta(\mathbf{r} - \mathbf{r}_{j}(t+t_{0}) + \mathbf{r}_{j}(t_{0})) \right\rangle,
\label{eq2/0}
\end{equation}
which measures the probability distribution for a particle displacement $\mathbf{r}$ in a time interval $t$. Since the present system is characterized by a translational symmetry along the nematic director, it is natural to separately study the diffusion along the \textit{z} axis and in the \textit{xy} plane. This can be done by partially integrating the SVHF on the $xy$ plane to get its longitudinal component

\begin{equation} 
 G_{s}^{\parallel}(z,t) = \frac{1}{N} \left\langle \sum_{j=1}^{N} \delta(z - z_{j}(t+t_{0}) + z_{j}(t_{0})) \right\rangle,
\label{eq2}
\end{equation}

and along the $z$ axis to get its transverse component, which can be further averaged over the azimuthal angle of $\mathbf{r}$
\begin{equation} 
 G_{s}^{\perp}(R,t) = \frac{1}{N} \left\langle \sum_{j=1}^{N} \delta(\mathbf{R} - \mathbf{R}_{j}(t+t_{0}) + \mathbf{R}_{j}(t_{0})) \right\rangle_{2 \pi}.
\label{eq3}
\end{equation}

In the above equations $(\mathbf{R}_j(t),z_j(t))$ is the position of particle $j$ at time $t$, $\delta$ is the Dirac delta, $\langle...\rangle$ stands for an ensemble average and the index $2 \pi$ indicates an additional average over the polar angle, which defines the bidimensional vector $\mathbf{R}=(x,y)$ with modulus $R=|\mathbf{R}|$. It should be noticed that for freely diffusive particles these functions are described by a Gaussian.

 \textit{\mdseries{Distinct-part of the van Hove function.}} A description of the influence of the surrounding particles background on the single particle diffusion is given by the DVHF, which is the probability distribution on the relative position $\mathbf{r}$ of two different particles at different times

\begin{equation} 
 G_{d}(\mathbf{r},t) = \frac{1}{N} \left\langle \sum_{j\neq i=1}^{N} \delta(\mathbf{r} - \mathbf{r}_{j}(t+t_{0}) + \mathbf{r}_{i}(t_{0})) \right\rangle.
\label{eq4}
\end{equation}
In order to separately study the caging regime related to the local fluid structure in the direction longitudinal and transverse to the nematic director, we actually measured the DVHF partially integrated over the size occupied by the spherocylinder in the \textit{xy} plane and along the \textit{z} axis, defined, respectively, by

\begin{equation} 
 G^{\parallel}_{d}(z,t) = \left(\frac {\pi D^{2}} {4}\right) ^{-1} \int_{0}^{D/2} R dR \int_{0}^{2 \pi} d\theta G_{d}(\mathbf{r},t), 
\label{eq5}
\end{equation}
\begin{equation} 
 G^{\perp}_{d}(R,t) = (2 \pi L)^{-1} \int_{-L/2}^{L/2} dz \int_{0}^{2 \pi} d\theta G_{d}(\mathbf{r},t),
\label{eq6}
\end{equation}
where we set $L=\min\{L_{1},L_{2}\}$ and $\theta$ is the azimuthal angle of $\mathbf{r}$ in the $xy$ plane.

 \textit{\mdseries{Non-Gaussian parameter.}} The deviations of the diffusion from Gaussian behavior can be estimated by the NGP, defined as \cite{rahman}

\begin{equation} 
 \alpha_{2}(t) = \frac{\langle \Delta \mathbf{r}^{4}(t) \rangle}{(1+2/d) \langle \Delta \mathbf{r}^{2}(t) \rangle^{2}} - 1,
\label{eq7}
\end{equation}
where $\Delta \mathbf{r}(t)$ is the displacement of a particle during a time interval $t$. The parameter $d$ corresponds to the number of dimensions considered, so that $d=1$ for the linear diffusion longitudinal to the nematic director ($\alpha_{2,z}(t)$) and $d=2$ for the planar transverse diffusion ($\alpha_{2,xy}(t)$). As pointed out in Ref. \cite{doliwa}, when treating mixtures one should be careful in not taking into account trivial non-Gaussianity due to a size-dependent particle mobility. In order to describe the dynamical heterogeneities related exclusively to the permanent barriers of the LC phase, one has to calculate the NGP $\alpha_{2}^{(i)}$ as defined in Eq. (\ref{eq7}) for each species $i=1,2$ separately and then perform an average weighted over the concentrations $x_i$, i.e.

\begin{equation} 
 \langle \alpha_{2}(t) \rangle = x_{1} \alpha_{2}^{(1)}(t) + x_{2} \alpha_{2}^{(2)}(t). 
\label{eq7bis}
\end{equation}
With these definitions heterogeneous diffusion can be detected when the NGP deviates from zero value.   

 \textit{\mdseries{Self-part of the intermediate scattering function.}} The structural relaxation of the system is conveniently described by measuring the self-part of the intermediate scattering function

\begin{equation} 
F_{s}(\mathbf{k},t) = \frac{1}{N} \left\langle \sum_{j=1}^{N} \exp[i \mathbf{k} \cdot ( \mathbf{r}_{j}(t+t_{0}) - \mathbf{r}_{j}(t_{0})) ] \right\rangle,
\label{eq8}
\end{equation}
which describes the density auto-correlations decay in the reciprocal space. Since most of the relevant structural information is contained at the first peak $\mathbf{k}^{*}$ of the structure factor, we can focus on the transverse and longitudinal relaxations by evaluating this function at $(k_{x}^{*},k_{y}^{*},0)$ and $(0,0,k_{z}^{*})$ respectively, so that $F_{s,xy}(t)=F_s((k_{x}^{*},k_{y}^{*},0),t)$ and $F_{s,z}(t)=F_s((0,0,k_{z}^{*}),t)$.

\section{Results}
\label{sec:Results}

\begin{figure}
\center
\includegraphics[scale=1.0]{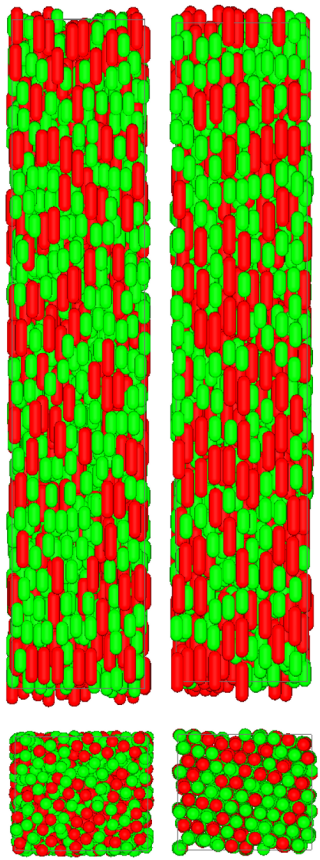}
\caption{\label{fig1} (Color online) Side and top views of two typical configurations in the nematic ($P^{*}=2.5$, left figure) and in the columnar phase ($P^{*}=4.0$, right figure) of a binary mixture of perfectly aligned hard spherocylinders with length-to-diameter ratios $L_1^*=2.1$ and $L_2^*=1.0$ and relative concentrations $x_1=0.375$ and $x_2=0.625$.}
\end{figure}

When the difference in length between the two components of a binary system of aligned hard spherocylinders is sufficiently high, the system undergoes a transition from a nematic to a columnar phase by increasing the pressure. The structure of the columnar phase is characterized by the development of an hexagonal order in the plane perpendicular to the nematic director \cite{stroobants}. This is the case for our parameter choice $L^{*}_1=2.1$ and $L^*_2=1.0$ as illustrated in Fig. \ref{fig1}, where two typical configurations in the nematic ($P^{*}=2.5$) and in the columnar phase ($P^{*}=4.0$) are compared.

\begin{figure}
\center
\includegraphics[scale=1.0]{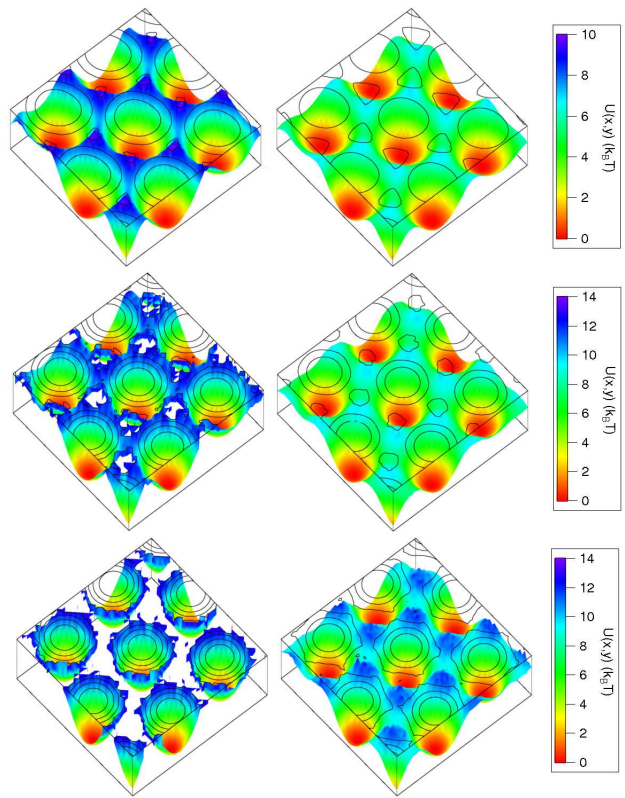}
\caption{\label{fig2} (Color online) Mean-field effective potential $U(x,y)$ in units of $k_{B}T$ in the bulk columnar phase of a binary mixture of perfectly aligned hard spherocylinders at $P^{*}=3.0$, $P^{*}=3.5$ and $P^{*}=4.0$ (from top to bottom). The images on the left correspond to the long rods (species $1$), whereas those on the right to the short ones (species $2$). In order to ease the visualization, the black lines at the top of each graph identify the isopotential points in the $xy$ plane with increments of $3 k_BT$.}
\end{figure} 

\begin{figure}
\center
\includegraphics[scale=1.0]{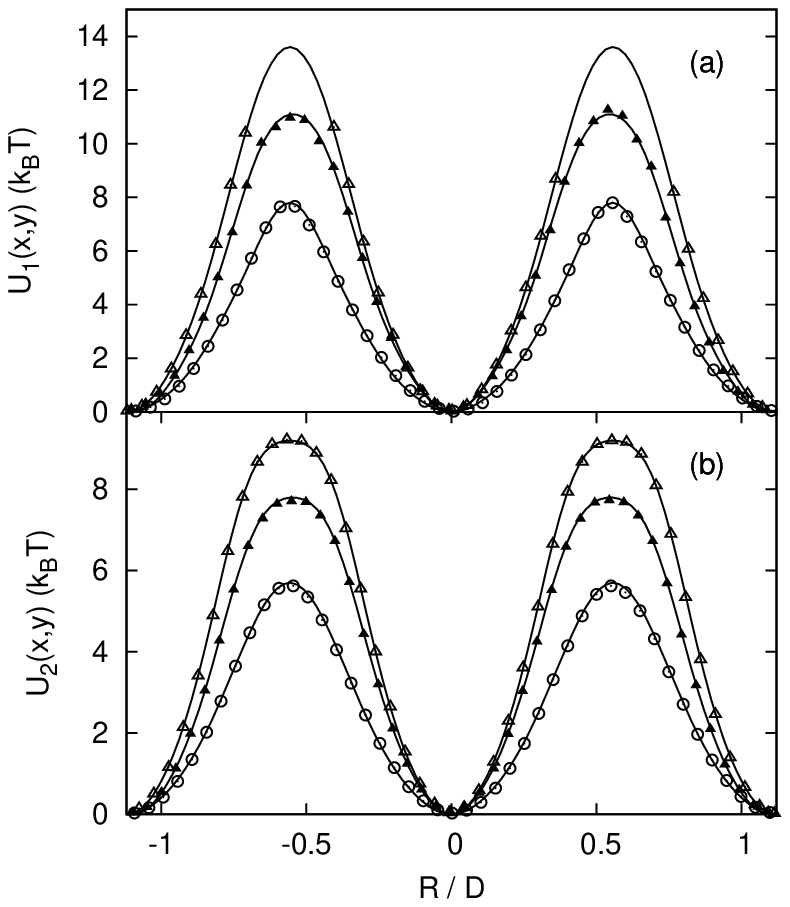}
\caption{\label{fig2bis} Transverse section of the mean-field effective potential in Fig. \ref{fig2} for a binary mixture of long (a) and short (b) hard spherocylinders at $P^*=3.0$ ($\circ$), $P^*=3.5$ ($\blacktriangle$) and $P^*=4.0$ ($\vartriangle$). The solid lines are fits (see text).}
\end{figure} 

The development of a long-range translational order allows to interpret the diffusion as the motion of a single particle subject to a periodic mean-field potential $U(x,y)$ as defined in Eq. (\ref{eq1}). This approach was applied in experiments \cite{lettinga} and simulations \cite{matena,patti,patti2} to characterize the hopping-type diffusion in smectic liquid crystals along the nematic director. These authors found that the free-energy cost for the layer-to-layer diffusion is in the order of few $k_{B}T$ per particle, depending mostly on the packing of the system, but also on the anisotropy and rotational degrees of freedom of the rods. Since the system studied here is composed of two species, the mean-field potential was evaluated separately for each component and is shown in Fig. \ref{fig2} for several pressures. The minima of the potential correspond to the lattice position, and the height of the energetic barriers gives a quantitative description of the energetic demand associated to a column-to-column jump. In order to estimate the height of the energetic barriers for each species as a function of pressure, we report in Fig. \ref{fig2bis} a transverse section of the energy landscapes in Fig. \ref{fig2}. Following the procedure in Ref. \cite{matena}, the experimental points in Fig. \ref{fig2bis} were fitted with a function 
\begin{equation} 
 U(R) = \sum_{k=1}^{n} U_k \left[ \sin \left( \frac {\pi R}{h} \right) \right]^{2k},
\label{pot_fit}
\end{equation}
with $U_k$ and $h$ fit parameters and $n=5$. As expected, the height of the potential barrier increases with the packing fraction and with the particle anisotropy, as already observed in Ref. \cite{patti}. At significant packing fractions, we detect that the energetic barriers appear higher for long rods. In particular, at $P^{*}=3.5$ and $4.0$ the column-to-column jumps become so rare that the associated statistics is too poor to furnish a precise estimate of the barrier height. In other words, the long rods are constrained to rattle in their cage, the jump to a neighboring column being too demanding. This is due to the fact that at high packing fraction no MC configuration showed a long rod in the region between the columns, with the result that the mean-field potential was characterized by an unphysical divergence. Furthermore, one should notice that the typical height of the barriers, close to and even higher than $10 k_{B}T$, is significantly higher than in the smectic phase \cite{matena,patti,patti2}. This can be seen by comparing our data at $P^{*}=3.5$ ($\eta=0.563$)
with those in Ref. \cite{matena} for the smectic phase of a system of aligned hard spherocylinders with $L^{*}=5.0$ at pressure $P^{*}=5.0$ ($\eta=0.563$). In the latter the height of the energy barrier reaches a value close to $8 k_{B}T$, which is expected to be even lower for shorter rods, as noticed in Ref. \cite{patti}.

\begin{figure}
\center
\includegraphics[scale=1.0]{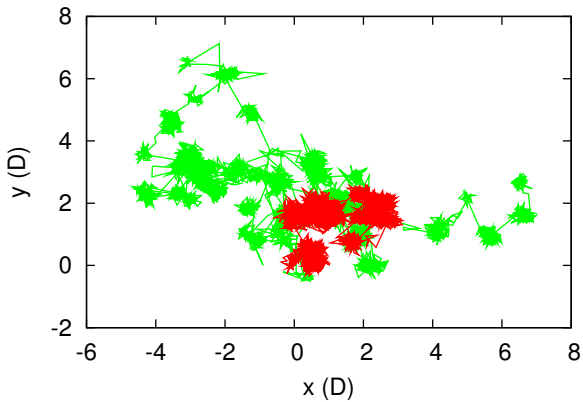}
\caption{\label{fig3} (Color online) A typical trajectory projected on the $xy$ plane of a long (red line) and short (green line) rod in the columnar phase at $P^{*}=3.0$ after an interval of time $\Delta t=380\tau$.} 
\end{figure}

The effect of the periodic mean-field potential can be further appreciated in Fig. \ref{fig3}, where we show a typical trajectory projected on the $xy$ plane of a long and a short particle at $P^{*}=3.0$. The difference with the Gaussian diffusion typical of a simple liquid, where the particle trajectories resemble the behavior of a random walker, is evident. In this case, the dynamics of the system is characterized by a hopping-type diffusion, in which each particle tends to rattle around the center of a column until it finds suitable conditions to overcome the energetic barrier and jump to another column in quasi-quantized steps. The spread in the total displacement between the two species is due to the higher barriers felt by the long rods, which significantly inhibit the inter-column diffusion. This behavior is observed in the whole range of pressures considered here, and its effects on the long-time relaxation dynamics of the system are crucial. More specifically, the long rods are expected to sample the configurational space on a time scale which might be significantly longer than that needed for the small ones. As a consequence, the decay of the correlation functions is strongly affected by the slow diffusion of the long particles, as we show later on.

\begin{figure}
\center
\includegraphics[scale=1.0]{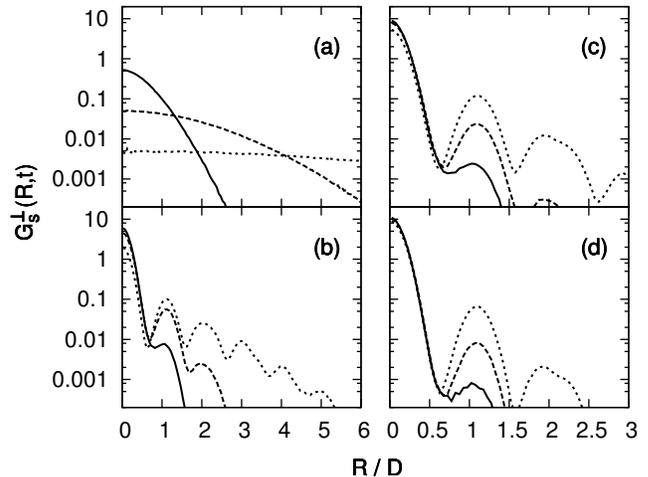}
\caption{\label{fig4} Transverse component of the self-part of the van Hove function $G_s^{\perp}(R,t)$ as a function of $R/D$ for a binary mixture of perfectly aligned hard spherocylinders with length-to-diameter ratio $L_1^*=2.1$ and $L_2^*=1.0$ and relative concentrations $x_1=0.375$ and $x_2=0.625$ at $t/\tau=1$ (solid lines), $t/\tau=10$ (dashed lines) and $t/\tau=100$ (dotted lines) for the system at pressure $P^{*}=2.5$ (a),  $P^{*}=3.0$ (b), $P^{*}=3.5$ (c) and $P^{*}=4.0$ (d).}
\end{figure}

\begin{figure}
\center
\includegraphics[scale=1.0]{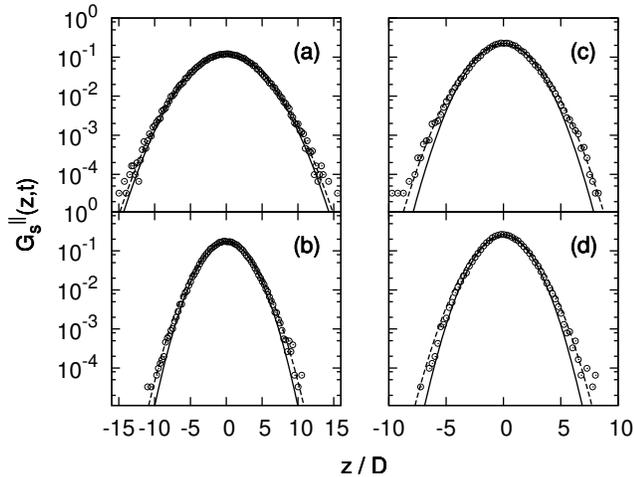}
\caption{\label{fig5} Longitudinal component of the self-part of the van Hove function $G_s^{\parallel}(z,t)$ as a function of $z/D$ for $t/\tau=20$ and pressure $P^{*}=2.5$ (a),  $P^{*}=3.0$ (b), $P^{*}=3.5$ (c) and $P^{*}=4.0$ (d). The curves refer to Gaussian fits over points near the origin (solid lines) and the tails (dashed lines).}
\end{figure}

In order to quantitatively study the distribution of displacements of the particles at different pressures and after different time intervals, we report in Fig. \ref{fig4} and \ref{fig5} the self-part of the van Hove function (SVHF) in its transverse and longitudinal components respectively. The comparison between the behavior of the transverse SVHF in the nematic (Fig. \ref{fig4}a) and in the columnar (Fig. \ref{fig4}b, \ref{fig4}c, and \ref{fig4}d) phase reveals a drastic change in the dynamics. The SVHF in the nematic phase is a monotonic function which broadens with time. By entering the columnar phase one observes the appearance of peaks which correspond to the positions of the hexagonal lattice in the $xy$ plane. As expected, after a fixed time interval the number and height of the peaks depend on the packing fraction of the system, so that by increasing the pressure the number of peaks decreases due to higher energetic barriers. These results confirm what was already observed for the smectic phase in experiments \cite{lettinga}, simulation \cite{patti} and theory \cite{bier}, i.e. the partial translational symmetry breaking in a liquid crystal gives rise to a non-Gaussian quasi-quantized diffusion related to a hopping-type dynamics. 

In Ref. \cite{bier} it was shown that, in order to accurately describe the dynamics of a liquid crystal system, it is not sufficient to take into account the permanent barriers due to the long-range structure, but also the transient caging effect given by the surrounding particles. In this sense, the local fluid structure can affect the diffusion by determining dynamical heterogeneities which make the system deviate from Gaussianity. A careful analysis on the longitudinal component of the SVHF in Fig. \ref{fig5} shows that this is indeed the case for the present columnar system. In fact, if along the $z$ axis the diffusion was Gaussian, it would be possible to fit the points in Fig. \ref{fig5} with a single Gaussian function. On the contrary, by performing this fit on different intervals on the $z$ axis, i.e. in the region near the origin (solid curve in figure) and the tails (dashed curve), one can observe that two different curves are obtained. Although in the present system the deviations between the two curves are small, this behavior manifests interesting resemblances with the heterogeneous dynamics of some amorphous systems, such as supercooled liquids and gels, where the two-Gaussian fitting is used to distinguish between ``slow'' and ``fast'' particles \cite{kegel,weeks}. In this sense, one can affirm that in the longitudinal direction the diffusion can be regarded as that of a dilute supercooled liquid, more than a normal liquid. The effect described so far should not be considered as strictly due to the columnar structure of the system, since analogous deviations are also observed in Fig. \ref{fig5}a for the nematic phase. Instead, the high packing fraction causes these small discrepancies from Gaussian diffusion. 

\begin{figure}
\center
\includegraphics[scale=1.0]{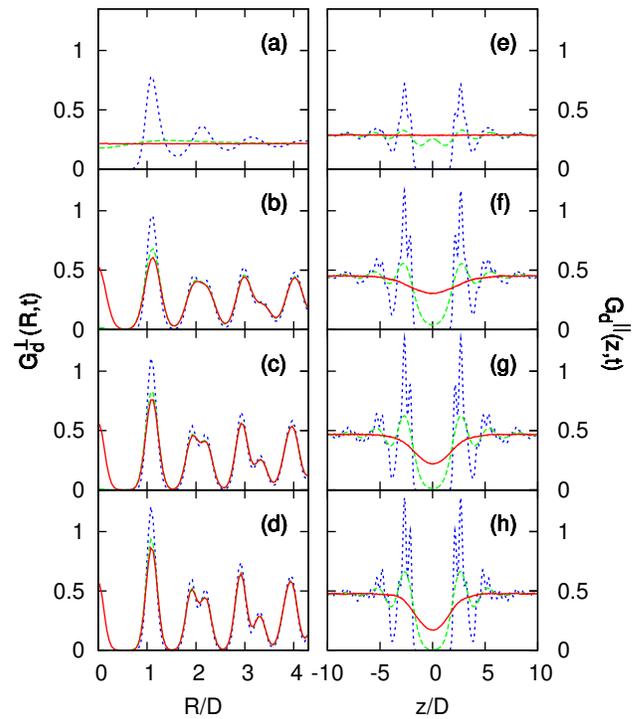}
\caption{\label{fig6} (Color online) Transverse $G_d^{\perp}(R,t)$ (a-d) and longitudinal $G_d^{\parallel}(z,t)$ (e-h) component of the distinct-part of the van Hove function evaluated for the same system and statepoints as in Fig. \ref{fig4} and \ref{fig5} and at $t/\tau=0.02$ (blue dotted line), $t/\tau=2$ (green dashed line), $t/\tau=20$ (red solid line) at $P^*=2.5$, $3.0$, $3.5$ and $4.0$ (from top to bottom).}
\end{figure}

A description of the transient caging regime due to the nearest-neighbor (solvation) shell around each particle can be given in terms of the distinct-part of the van Hove function (DVHF) defined in Eq. (\ref{eq5}) and (\ref{eq6}) and reported in Fig. \ref{fig6} for $t/\tau=0.02$, $2$ and $20$. According to the definition given in Eq. (\ref{eq4}), at time $t=0$ the DVHF coincides with the pair distribution function, and it is thus characterized by a region around the origin where its value is equal to zero due to the excluded volume interaction. On the other hand, in the limit $t \rightarrow \infty$ the DVHF is expected to be a constant in a translationally homogeneous system due to the decay of the positional correlations; this is not the case in presence of translational order, since the mutual position of two particles at different times is influenced by the permanent long-range structure of the whole system. At $t/\tau=0.02$ a region around the origin where the DVHF is close to zero suggests that each particle is still rattling around its initial position. Beyond this region a series of peaks indicate the preferential positions of the particles with respect to the one placed at the origin at the initial time. In the nematic phase at $t/\tau=0.02$ one can recognize in both the transverse (Fig. \ref{fig6}a) and longitudinal (Fig. \ref{fig6}e) components the liquid-like structure of the system, where the lack of long-range order is testified by the rapid decay of the peaks by moving away from the origin. One should also notice the speed at which the gap region around the origin is filled, giving rise to a DVHF almost constant already at $t/\tau=2$. Therefore, during this time interval a given particle $i$ can escape the trapping cage formed by its nearest neighbors $j$, and the space originally occupied by $i$ will be filled by one of the $j$ particles. The situation changes appreciably when we pass to the columnar phase, where the long-range modulations in the transverse component of the DVHF (Fig. \ref{fig6}b-d) indicate the presence of a permanent structure. On the other hand, the longitudinal component (Fig. \ref{fig6}f-h) does not display any dramatic change in shape, but from its time evolution one can observe that the time a particle needs to leave its initial position increases considerably. In fact, whereas the relaxation times of the longitudinal DVHF in the three systems manifesting columnar order are comparable (Fig. \ref{fig6}f-h), one can notice a faster relaxation in the nematic phase (Fig. \ref{fig6}e), which cannot be due to the difference in packing fraction exclusively. This seems to suggest that the inhomogeneous structure and the resulting dynamics in the transverse plane appreciably affect the dynamics in the longitudinal direction. We argue that the higher in-plane mobility of the nematic phase with respect to the columnar affects the mobility along the nematic director albeit only slightly. This is coherent with the results of Ref. \cite{bier}, where a coupling between transverse and longitudinal diffusion in the smectic phase was pointed out.    

\begin{figure}
\includegraphics[scale=1.0]{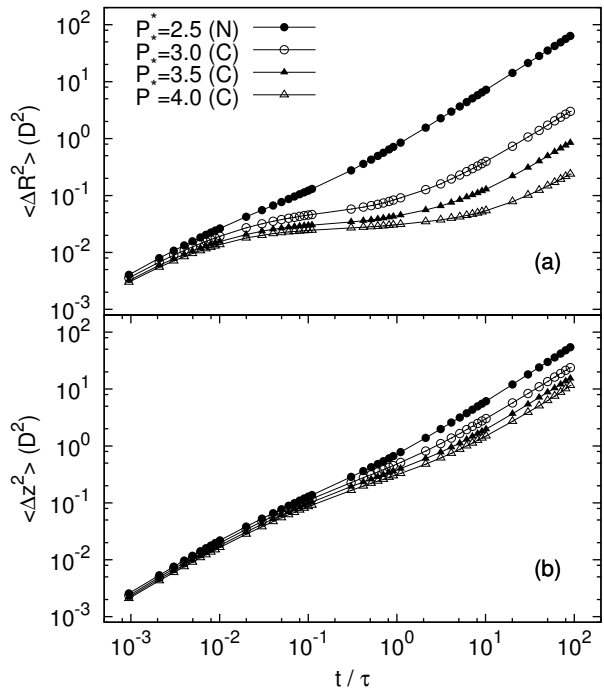}
\caption{\label{fig7} Mean square displacement in the direction perpendicular $\langle \Delta R^2 \rangle$ (a) and parallel $\langle \Delta z^2 \rangle$ (b) to the nematic director in units of $D^2$ as a function of $t/\tau$ for the same system and statepoints as in Fig. \ref{fig4}-\ref{fig6}.}
\end{figure}

\begin{figure}
\includegraphics[scale=1.0]{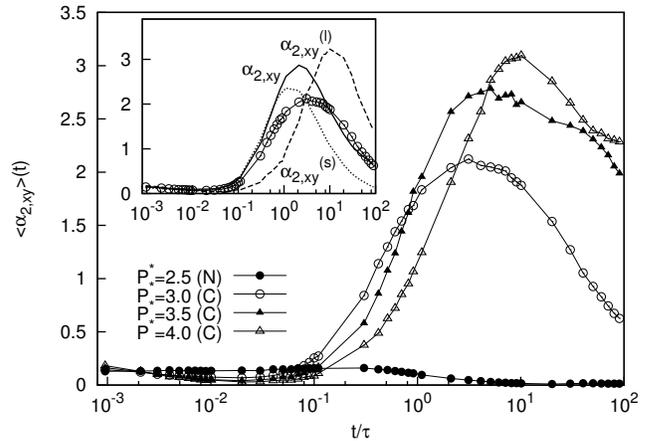}
\caption{\label{fig7bis} Non-Gaussian parameter $\langle \alpha_{2,xy} \rangle (t)$ of the diffusion in the plane perpendicular to the nematic director as a function of $t/\tau$ for the same system and statepoints as in Fig. \ref{fig4}-\ref{fig7}. In the inset we compare at $P^{*}=3.0$ the non-Gaussian parameter averaged over the concentrations (open circles) with those relative to species $1$ (dashed line), to species $2$ (dotted line) and of the whole system (solid line).}
\end{figure}

An alternative way to analyze dynamical heterogeneities is to look for deviations from linearity of the mean square displacement. The effect of local cage trapping in systems close to dynamical arrest and the presence of permanent long-range inhomogeneities as in liquid crystals manifest themselves in a region at intermediate times where the dynamics is strongly sub-diffusive. In Fig. \ref{fig7} we show the MSD both in the $xy$ plane and in the $z$ direction. In the plane perpendicular to the nematic director (Fig. \ref{fig7}a) one can appreciate the almost linear trend of the MSD in the nematic phase, whereas by increasing the pressure and going to the columnar phase a plateau region appears, manifesting the development of a heterogeneous dynamics. These deviations from linearity are tightly related to the non-Gaussian behavior of the self-part of the van Hove function, and can be quantitatively estimated by the non-Gaussian parameter defined in Eq. (\ref{eq7}). In Fig. \ref{fig7bis} we report the NGP in the $xy$ plane averaged over the species concentrations as described in Eq. (\ref{eq7bis}). This parameter remains close to zero in the nematic phase, while it displays a peak at intermediate times in the columnar phase indicating deviations from Gaussianity. On the other hand, along the $z$ direction (not shown here) the NGP does not deviate significantly from zero. The choice of calculating the NGP for the whole system by averaging over the value it assumes for the two species separately allows to take into account just the effects related to the long-range structure of the system. For the sake of completeness we show in the inset of Fig. \ref{fig7bis} a comparison between the NGP at $P^*=3.0$ for each species, their weighted average and that corresponding to the whole system. As expected, the operation of average does not affect significantly the position of the peak, but decreases only the peak height, suggesting that in this way the non-Gaussianity due to particle size difference is subtracted. One should notice that this particular treatment is not necessary for the rest of the physical properties measured in this paper, since with the exception of the NGP they result to be linear in the particle species, i.e. their value for the system as a whole corresponds to a weighted average over the species.

The dynamic inhomogeneities as captured by both the MSD and the NGP allow to identify three different time intervals. At short times the MSD follows the usual linear trend and the NGP maintains a value close to zero, which means that particles diffuse freely since they do not feel yet the trapping cage due to the surrounding particles. At intermediate times the MSD becomes strongly sub-diffusive and the NGP is characterized by a monotonic growth, thus meaning that the free diffusion is inhibited by the columnar structure of the fluid. At this stage one can distinguish between particles which still rattle around the position of their column and others which succeeded in overcoming the energetic barrier and jumped to another column. The end of the subdiffusive plateau and the return to the linear trend of the MSD roughly correspond to the peak of the NGP, which starts decreasing monotonically to zero indicating the end of the caging regime, i.e. most of the particles succeeded in leaving their initial column. A deeper inspection on the pressure dependence of the NGP shows that the degree of non-Gaussianity, i.e. the height of the peak, and the duration of the caging regime, i.e. the position of the peak, increase with packing fraction. This fact can be explained by considering that the cage escape is related to a rearrangement of the surrounding particles, which becomes slower at higher packing fraction as it involves more of them. Furthermore, the small deviations from linearity in the MSD in the direction parallel to the nematic director confirm the presence of a weakly heterogeneous dynamics, as already pointed out by analyzing the self-part of the van Hove function in this direction.

\begin{figure}
\center
\includegraphics[scale=1.0] {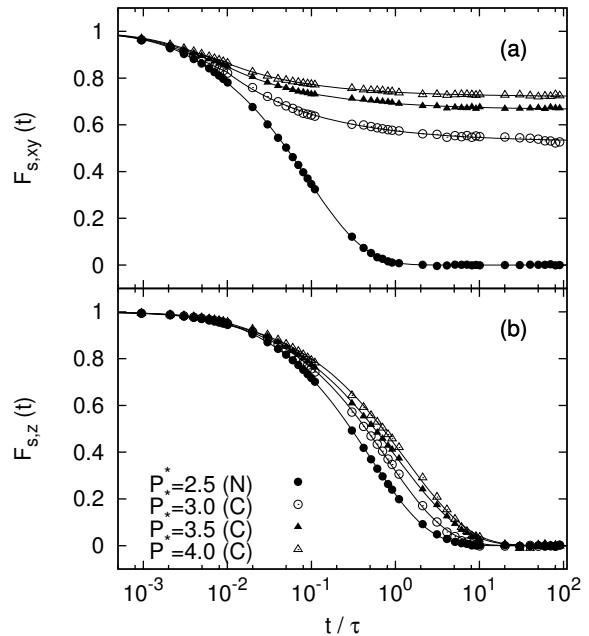}
\caption{\label{fig8} Self-part of the intermediate scattering function $F_{s,xy}(t)$ and $F_{s,z}(t)$ evaluated at wave vectors corresponding to the first peak of the structure factor in the plane perpendicular to the nematic director (a) and along it (b). The data correspond to the same system and statepoints as in Fig. \ref{fig4}-\ref{fig7bis}. The solid lines are fits (see text).}
\end{figure}

Finally, the structural relaxation of the system is analyzed in terms of the self-part of the intermediate scattering function (SISF) defined in Eq. (\ref{eq8}). Whereas along the $z$ direction the relaxation is characterized by a single step decay at each pressure (Fig. \ref{fig8}b), a plateau region, which characterizes the relaxation in the $xy$ plane at intermediate times, develops in the columnar phase. This plateau, whose value increases with pressure, indicates the time extension of the cage regime and is expected to divide a short-time decay ($\beta$-relaxation) from a long-time one ($\alpha$-relaxation). As previously observed in recent work on smectic liquid crystals \cite{matena,patti,patti2} and in out-of-equilibrium supercooled liquids \cite{berthier}, the SISF decays likely to zero at long times, indicating the loss of density auto-correlations. This kind of behavior was described for the smectic phase in Refs. \cite{matena,patti,patti2}, where the $\alpha$-relaxation decay was fitted by a stretched exponential function of the form $\exp[(t/t_{r})^{\beta}]$ with $\beta\simeq0.6$ and $t_r$ the characteristic relaxation time. In the present simulations we did not observe any $\alpha$-relaxation as the relaxation time of the systems exceeds probably our simulation time. On the other hand, from the data available a close accordance with the features of the structural relaxation of the smectic phase can be observed. In particular, the $\beta$-relaxation in the $xy$ plane is reasonably described by an exponential decay, as expected for simple liquids, due to the lack of interactions of the particles with the nearest neighbors at small times. Also the relaxation along the $z$ axis resembles accurately the relaxation of the smectic phase inside the smectic layers. In fact, in both these cases the SISF depends weakly on the pressure and it is characterized by an exponential decay at small times, which eventually becomes a stretched exponential with $\beta\simeq0.6$. In this sense we can confirm what the authors observed in Ref. \cite{matena}, i.e. the relaxation of a liquid crystal in the direction(s) in which the system is homogeneous is closer to that of a low-density supercooled liquid, than a simple liquid, where instead an exponential relaxation is to be expected.

\section{Conclusions}
\label{sec:Conclusions}

In summary, we used Monte Carlo simulations to analyze for the first time the presence of dynamical heterogeneities in a columnar liquid crystal of perfectly aligned hard spherocylinders. The long-range hexagonal order in the plane perpendicular to the nematic director determines an effective mean-field potential, whose effect is to maintain particles inside a column preventing them to occupy a position in between the columns. In analogy with previous analyses on the smectic phase, the height of the energetic barriers of this effective potential increases with the packing fraction and the particle anisotropy. As a consequence, in the $xy$ plane the dynamics of a rod is characterized by a quasi-quantized behavior in which particles rattle around the position of the column and jump to another column only when the configuration of the surrounding particles allows it.
 
The rattling-and-jumping dynamics in the in-plane evolution of the system gives rise to three different time regimes. At very short times, the particles diffuse almost freely because they do not feel yet the presence of the trapping cage formed by their surrounding nearest neighbors. At this stage the behavior of the system is that typical of a simple fluid, characterized by a Gaussian distribution of displacements, a linear mean square displacement and a fast exponential structural relaxation. A second stage starts when particles begin experiencing the cage due to the long-range structure of the system, in such a way that the diffusion results to be inhibited and only occasionally a column-to-column jump takes place and is made possible by the instantaneous configuration of the system. As a result, the mean square displacement as well as the self-intermediate scattering function develop a plateau, which testifies the slowing down of the dynamics and whose time extension increases with packing fraction. On the other hand, the distribution of displacements show marked deviations from Gaussianity due to the appearance of peaks which correspond to the lattice positions in the plane. Nonetheless, after longer time intervals the number of ``fast'' particles, which succeeded in overcoming the energetic barrier, increases with respect to the ``slow'' ones. Consequently, when most of the particles succeeded in leaving their initial column, a second diffusive regime starts, indicating the end of the cage regime.

We observed interesting analogies with the dynamics in smectic phases by considering the in-column dynamics. In fact, along the direction defined by the nematic director, the system does not develop any long-range order, and it is thus expected to behave like a liquid. On the other hand, we noticed interesting, although slight, deviations from Gaussian diffusion both in the distribution of displacements and in the mean square displacement. As far as the structural relaxation is concerned, this fact is testified by a self-intermediate scattering function well approximated by a stretched-exponential, as it happens in dense liquids. In this sense we confirm previous studies on the smectic phase, that along the direction in which a liquid crystal does not develop any long-range order the dynamics is similar to a dense liquid. These results are to be compared with recent experiments on the columnar phase of a suspension of \textit{fd} virus particles \cite{pouget}, where huge discrepancies from Gaussianity were observed along the nematic director. We argue that the higher length-to-diameter ratio, the flexibility or the charge of the rods could account for a more pronounced non-Gaussian diffusive behavior than what was observed in the present study.

\section{Acknowledgements}
\label{sec:Acknowledgements}

It is a pleasure to thank M. P. Lettinga, E. Grelet and E. Pouget for sharing preliminary experimental results that motivated this work and D. El Masri and P. van der Schoot for stimulating discussions. 

This work is financed by a NWO-VICI grant and is part of the research program of the ``Stichting door Fundamenteel Onderzoek der Materie (FOM)'', which is financially supported by the ``Nederlandse organisatie voor Wetenschappelijk Onderzoek (NWO)''.


\begin{thebibliography} {1}

\bibitem{onsager}
L. Onsager, \textit{Ann. N. Y. Acad. Sci.} {\bf 51}, 627 (1949).

\bibitem{israelachvili}
J. N. Israelachvili and B. W. Ninham, \textit{J. Colloid Interface Sci.} {\bf 58}, 14 (1977).

\bibitem{forsyth}
P. A. Forsyth et al., \textit{Adv. Colloid Interface Sci.} {\bf 9}, 37 (1978).

\bibitem{flory}
P. J. Flory, \textit{Macromolecules} {\bf 11}, 1138 (1978).

\bibitem{frenkel2}
D. Frenkel et al., \textit{Nature} {\bf 332}, 822 (1998).

\bibitem{stroobants1}
A. Stroobants, H. N. W. Lekkerkerker, D. Frenkel, \textit{Phys. Rev. Lett.} {\bf 57}, 1452 (1986).

\bibitem{stroobants2}
A. Stroobants, H. N. W. Lekkerkerker, D. Frenkel, \textit{Phys. Rev. A} {\bf 36}, 2929 (1987).

\bibitem{veerman}
J. A. C. Veerman and D. Frenkel, \textit{Phys. Rev. A} {\bf 43}, 4334 (1991).

\bibitem{bolhuis}
P. Bolhuis and D. Frenkel, \textit{J. Chem. Phys.} \textbf{106}, 666 (1997).

\bibitem{stroobants}
A. Stroobants, \textit{Phys. Rev. Lett.} \textbf{69}, 2388 (1992).

\bibitem{cui}
S.-M. Cui and Z. Y. Chen, \textit{Phys. Rev. E} \textbf{50}, 3747 (1994).

\bibitem{bates}
M. A. Bates and D. Frenkel, \textit{J. Chem. Phys.} \textbf{109}, 6193 (1998).

\bibitem{martinez}
Y. Mart\'{i}nez-Rat\'{o}n, E. Velasco, L. Mederos, \textit{J. Chem. Phys.} \textbf{123}, 104906 (2005).

\bibitem{varga}
S. Varga et al., \textit{Mol. Phys.} \textbf{107}, 2481 (2009).

\bibitem{frenkel3}
D. Frenkel, \textit{Liquid Crystals} \textbf{5}, 929 (1989).

\bibitem{veerman2}
J. A. C. Veerman and D. Frenkel, \textit{Phys. Rev. A} {\bf 45}, 5632 (1992).

\bibitem{cuetos}
A. Cuetos and B. Martinez-Haya, \textit{J. Chem. Phys.} \textbf{129}, 214706 (2008).

\bibitem{filion}
L. Filion, M. Marechal et al., \textit{Phys. Rev. Lett.} {\bf 103}, 188302 (2009).

\bibitem{bohle}
A. M. Bohle et al., \textit{Phys. Rev. Lett.} \textbf{76}, 1396 (1996).

\bibitem{wensink}
H. H. Wensink, \textit{J. Chem. Phys.} \textbf{126}, 194901 (2007).

\bibitem{hentschke}
R. Hentschke and J. Herzfeld, \textit{Phys. Rev. A} \textbf{44}, 1148 (1991).

\bibitem{sear}
R. P. Sear, \textit{Phys. Rev. E} \textbf{55}, 5820 (1997).

\bibitem{cinacchi}
G. Cinacchi and L. De Gaetani, \textit{Phys. Rev. E} \textbf{77}, 051707 (2008).

\bibitem{grelet2}
E. Grelet, \textit{Phys. Rev. Lett.} \textbf{100}, 168301 (2008).

\bibitem{hess}
S. Hess, D. Frenkel, M. P. Allen, \textit{Mol. Phys.} \textbf{74}, 765 (1991).

\bibitem{lowen}
H. L\"{o}wen, \textit{Phys. Rev. E} \textbf{59}, 1989 (1999).

\bibitem{bates2}
M. A. Bates and G. R. Luckhurst, \textit{J. Chem. Phys.} \textbf{120}, 394 (2004).

\bibitem{cifelli}
M. Cifelli, G. Cinacchi, L. De Gaetani, \textit{J. Chem. Phys.} \textbf{125}, 164912 (2006).

\bibitem{bu}
Z. Bu et al., \textit{Macromolecules} \textbf{27}, 6871 (1994).

\bibitem{vbruggen1}
M. P. B. van Bruggen, H. N. W. Lekkerkerker, J. K. G. Dhont, \textit{Phys. Rev. E} \textbf{56}, 4394 (1997).

\bibitem{vbruggen2}
M. P. B. van Bruggen et al., \textit{Phys. Rev. E} \textbf{58}, 7668 (1998).

\bibitem{cush}
R. C. Cush and P. S. Russo, \textit{Macromolecules} \textbf{23}, 8659 (2002).

\bibitem{lettinga2}
M. P. Lettinga, E. Barry, Z. Dogic, \textit{Europhys. Lett.} \textbf{71}, 692 (2005).

\bibitem{lettinga}
M. P. Lettinga and E. Grelet, \textit{Phys. Rev. Lett.} {\bf 99}, 197802 (2007).

\bibitem{kob}
W. Kob et al., \textit{Phys. Rev. Lett.} {\bf 79}, 2827 (1997).

\bibitem{gray}
A. Gray-Weale, P. A. Madden, \textit{J. Phys. Chem. B} {\bf 108}, 6624 (2004).

\bibitem{puertas}
A. M. Puertas, M. Fuchs, M. E. Cates, \textit{J. Chem. Phys.} {\bf 121}, 2813 (2004).

\bibitem{gao}
Y. Gao, M. L. Kilfoil, \textit{Phys. Rev. E} {\bf 79}, 051406 (2009).

\bibitem{bier}
M. Bier, R. van Roij, M. Dijkstra, and P. van der Schoot, \textit{Phys. Rev. Lett.} {\bf 101}, 215901 (2008).

\bibitem{grelet}
E. Grelet et al., \textit{J. Phys.: Condens. Matter} {\bf 20}, 494213 (2008).

\bibitem{matena}
R. Matena, A. Patti, M. Dijkstra, \textit{Phys. Rev. E} \textbf{81}, 021704 (2010).

\bibitem{patti}
A. Patti, D. El Masri, R. van Roij, and M. Dijkstra, \textit{J. Chem. Phys.} {\bf 132}, 224907 (2010).

\bibitem{patti2}
A. Patti, D. El Masri, R. van Roij, and M. Dijkstra, \textit{Phys. Rev. Lett.} \textbf{103}, 248304 (2009).

\bibitem{pouget}
E. Pouget, E. Grelet, M. P. Lettinga, \textit{unpublished}.

\bibitem{doliwa2}
B. Doliwa and A. Heuer, \textit{Phys. Rev. Lett.} \textbf{80}, 4915 (1998).

\bibitem{doliwa}
B. Doliwa and A. Heuer, \textit{J. Phys.: Condens. Matter} {\bf 11}, A277 (1999).

\bibitem{berthier}
L. Berthier and W. Kob, \textit{J. Phys.: Condens. Matter} \textbf{19}, 205130 (2007).

\bibitem{brambilla}
G. Brambilla et al., \textit{Phys. Rev. Lett.} \textbf{102}, 085703 (2009).

\bibitem{berthier2}
L. Berthier and T. A. Witten, \textit{Phys. Rev. E} \textbf{80}, 021502 (2009).

\bibitem{huitema}
H. E. A. Huitema and J. P. van der Eerden, \textit{J. Chem. Phys.} \textbf{110}, 3267 (1999).

\bibitem{frenkel}
D. Frenkel and B. Smit, \textit{Understanding Molecular Simulations: From Algorithms to Applications}, second ed., Academic Press, San Diego, 2001.

\bibitem{pryamitsyn}
V. Pryamitsyn and V. Ganesan, \textit{J. Chem. Phys.} \textbf{128}, 134901 (2008).

\bibitem{sanz}
E. Sanz and D. Marenduzzo, \textit{J. Chem. Phys.} \textbf{132}, 194102 (2010).

\bibitem{tirado}
M. M. Tirado and G. C. de la Torre, \textit{J. Chem. Phys.} \textbf{71}, 2581 (1979).

\bibitem{tirado1}
M. M. Tirado and G. C. de la Torre, \textit{J. Chem. Phys.} \textbf{81}, 2047 (1984).

\bibitem{vanhove}
L. Van Hove, \textit{Phys. Rev.} {\bf 95}, 249 (1954).

\bibitem{rahman}
A. Rahman, \textit{Phys. Rev.} {\bf 136}, A405 (1964).

\bibitem{kegel}
W. K. Kegel and A. van Blaaderen, \textit{Science} \textbf{287}, 5451 (2000).

\bibitem{weeks}
E. R. Weeks and D. A. Weitz, \textit{Chem. Phys.} \textbf{284}, 361 (2002).

\end{thebibliography}
\end{document}